\documentclass[10pt,conference]{IEEEtran}

\usepackage[colorlinks=true, urlcolor=blue, linkcolor=black, citecolor=black]{hyperref} % Added for clickable URLs, hidelinks removes boxes

\usepackage{amsmath,amssymb,amsfonts}
\usepackage{algorithmic} % For algorithms
\usepackage{graphicx}
\usepackage{textcomp}
\usepackage{xcolor}
\usepackage{url} % hyperref also handles \url
\usepackage{svg}
\usepackage{comment}
\usepackage[breakable]{tcolorbox}
\usepackage{threeparttable}
\usepackage{booktabs}
\usepackage{array, makecell}
\usepackage{cite} % For IEEE style citations like [1]-[3]
\usepackage[switch]{lineno}

\begin{document}
\title{Agents in the Sandbox: End-to-End Crash Bug Reproduction for Minecraft}
\IEEEspecialpapernotice{Accepted to ASE 2025 — 
\href{https://conf.researchr.org/details/ase-2025/ase-2025-papers/244/Agents-in-the-Sandbox-End-to-End-Crash-Bug-Reproduction-for-Minecraft}{ASE 2025 paper page}}
\author{
\IEEEauthorblockN{Eray Yapağcı}
\IEEEauthorblockA{Electrical and Electronics
Engineering Department\\
Bilkent University\\
Ankara, Turkey\\
eray.yapagci@bilkent.edu.tr}
\and
\IEEEauthorblockN{Yavuz Alp Sencer Öztürk}
\IEEEauthorblockA{Computer Engineering Department\\
Bilkent University\\
Ankara, Turkey\\
alp.ozturk@ug.bilkent.edu.tr}
\and
\IEEEauthorblockN{Eray Tüzün}
\IEEEauthorblockA{Computer Engineering Department\\
Bilkent University\\
Ankara, Turkey\\
eraytuzun@cs.bilkent.edu.tr}
}

\maketitle

\begin{abstract}
Reproducing game bugs, particularly crash bugs in continuously evolving games like Minecraft, is a notoriously manual, time-consuming, and challenging process to automate; insights from a key decision maker from Minecraft we interviewed confirm this, highlighting that a substantial portion of crash reports necessitate manual scenario reconstruction. Despite the success of LLM-driven bug reproduction in other software domains, games, with their complex interactive environments, remain largely unaddressed. This paper introduces BugCraft, a novel end-to-end framework designed to automate the reproduction of crash bugs in Minecraft directly from user-submitted bug reports, addressing the critical gap in automated game bug reproduction. BugCraft employs a two-stage approach: first, a Step Synthesizer leverages LLMs and Minecraft Wiki knowledge to transform bug reports into high-quality, structured steps to reproduce (S2R). Second, an Action Model, powered by a vision-based LLM agent and a custom macro API, executes these S2R steps within Minecraft to trigger the reported crash. To facilitate evaluation, we introduce BugCraft-Bench, a curated dataset of Minecraft crash bug reports. On BugCraft-Bench, our framework end-to-end reproduced 34.9\% of crash bugs with GPT-4.1, outperforming baseline computer-use models by 37\%. BugCraft demonstrates the feasibility of automated reproduction of crash bugs in complex game environments using LLMs, opening promising avenues for game testing and development. 
%The framework and the BugCraft-Bench dataset pave the way for future research in automated game bug analysis and hold potential for generalization to other interactive game platforms.
Finally, we make our code open at \url{https://bugcraft2025.github.io}.
\end{abstract}


\section{Introduction} \label{sec:introduction}

In the maintenance phase of the game development cycle, it is crucial to provide updates that improve the game to maintain player retention and enhance engagement \cite{Zhong_Xu_2021}. These game updates should also fix bugs for improved user experience. As a type of bug, crash bugs are especially severe because they prevent the user from experiencing the game. They are also difficult to fix, being the most recurring bug type even after updates addressing the bug \cite{Truelove_Santana_De_Almeida_Ahmed_2021}.

A primary source of information for these updates is the user-submitted bug reports. A prime example is Mojira\footnote{\url{https://bugs.mojang.com/issues}}, the public bug tracker for Minecraft developer Mojang Studios. Mojira hosts more than 12 years of bugs for various versions of Minecraft\footnote{\url{https://bugs.mojang.com/browse/MC-1}}. However, many of these reports are of low quality, often missing essential components like S2R, expected outcomes, and can have mistakes in them \cite{Zimmermann_Premraj_Bettenburg_Just_Schroter_Weiss_2010, Soltani_Hermans_Bäck_2020,bugTrackingProcessSmells,Altun2025ProcessSmells,QAMAR2022106972}. These difficulties prolong bug lifecycles \cite{AnalyzingBugLifeCycles}, as challenging reproduction decreases reviewer motivation and delays confirmation. The studies showing low-quality bug reports across numerous projects suggest that Mojira could potentially face similar challenges.

Due to the excessive amount of labor involved, game testing is an entire field in itself, and it is a notoriously manual process that takes a significant amount of time \cite{Politowski_Petrillo_Gueheneuc_2021}. To confirm the real-world impact of this challenge, we conducted a semi-structured interview with a key decision maker from Minecraft in May 2025. This expert emphasized the critical importance of successful reproduction, noting that once a bug is reproducible, a significant portion of the diagnostic challenge is considered overcome, bringing considerable relief to the development team. The interview confirmed that their teams manually reconstruct scenarios for a significant volume of crash reports, highlighting the substantial manual effort involved and the potential utility of automated solutions like BugCraft. This critical need for automation, especially in a huge sandbox game, motivates our work. Most games lack automated test frameworks because implementing and generalizing such systems is challenging. For example, Minecraft's internal unit test framework GameTest\footnote{\url{https://www.minecraft.net/en-us/creator/article/get-started-gametest-framework}} lacks the support to reproduce UI movement and precise player interactions, only covering a part of what is a huge sandbox game. Moreover, developers need to write unit tests manually for each specific bug to be integrated into the game.

Looking at a more general usage, automated bug reproduction systems have been developed \cite{Kang_Yoon_Yoo_2023}. With the advent of large language models (LLMs) that are capable of tool calling (i.e., interacting with external tools/APIs) \cite{qin2024toollearningfoundationmodels} and code generation \cite{jiang2024surveylargelanguagemodels}, the implementations of these systems have become more general. Similar works \cite{Liu_Li_Chen_Wang_Wu_Wang_Hu_Wang_2024, Wang_Zhao_Feng_Zhang_Halfond, Feng_Chen_2024} exist for reproducing crash bugs in Android apps through interacting with the available GUI. These works are limited in interacting with the game, which needs action-taking in a complex environment. To our knowledge, there are no works for end-to-end automated crash bug reproduction in an open-ended game environment like Minecraft.

In this paper, we present BugCraft, an agentic bug reproduction framework for Minecraft, which makes the following novel contributions:
\begin{itemize}
    \item The first end-to-end framework for automated crash bug reproduction in a complex, 3D game environment. BugCraft tackles the full pipeline from an unstructured bug report to the generation of a script and a final log that can reliably reproduce the crash.

    \item A novel, multi-stage Step Synthesizer that transforms ambiguous bug reports into actionable plans by synthesizing information from diverse sources, including user comments and domain-specific knowledge bases. This is achieved through a process of RAG, intelligent analysis of the game state to find reliable solutions, and a final structuring phase using what we term Step Clusters, granting the agent crucial flexibility during execution.

    \item BugCraft-Bench, the first publicly available benchmark for this task, containing 86 manually confirmed and reproducible crash bug reports to facilitate future research.
\end{itemize}
We evaluate BugCraft against competitive baselines including OpenAI's Computer Use Agent (OpenAI-CUA) and UI-TARS-1.5-7B  \cite{qin2025ui}, demonstrating a 34.9\% success rate that outperforms existing approaches. Additionally, we test BugCraft with multiple LLMs (GPT-4o and GPT-4.1) to demonstrate how model improvements directly enhance framework performance. To ensure full reproducibility, we provide the BugCraft-Bench, the source code, run logs, and annotations at our project website: \url{https://bugcraft2025.github.io}.

\begin{figure}[h]
    \centering
    \includegraphics[width=1\linewidth]{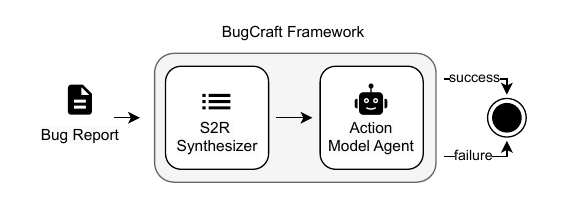}
    \caption{The BugCraft framework, illustrating the two-stage process of S2R synthesis and action model execution.}
    \label{fig:intro_diagram}
\end{figure}

%The key results of our framework's benchmark on our dataset are as follows: We were able to reproduce 31.7\% of the crash bug reports and generate accurate plans for reproduction with 66.3\% accuracy.

% Our paper begins by discussing related work in Section~\ref{sec:related_work}, followed by an overview of the framework in Section~\ref{sec:methodology}. We define our research questions in Section~\ref{sec:study}, evaluate the framework in Section~\ref{sec:results}, and discuss our findings in Section~\ref{sec:discussion}. The paper mentions the threats to validity in Section~\ref{sec:threats_to_validity},  and concludes with a summary of contributions in Section~\ref{sec:conclusion}.

\section{Related Work} \label{sec:related_work}

% This section provides an overview of related topics, focusing on bug reproduction in games and the development of agents powered by LLMs. We also discuss similar research efforts relevant to our framework.

\subsection{Bug Reproduction in Games}

Game testing is traditionally a manual and time-consuming process, and the primary approach involves playtesting, in which testers manually play the game to identify issues \cite{Politowski_Petrillo_Gueheneuc_2021}. Automated unit testing is often challenging for game developers due to the dynamic nature of game code and the complexity involved in capturing game states programmatically \cite{Pascarella_Palomba_Di_Penta_Bacchelli_2018}. In continuous development cycles, it becomes essential for developers to address existing bugs through regular updates to improve the gameplay experience. However, certain types of bugs, particularly crash bugs, are prone to reoccurrence even after an update is released to fix them. One developer in a recent survey attributed this issue to limited or non-available testing time, highlighting the constraints of manual testing in the fast-paced, deadline-driven environment typical of game development \cite{Truelove_Santana_De_Almeida_Ahmed_2021}.

Wuji \cite{Zheng_Xie_Su_Ma_Hao_Meng_Liu_Shen_Chen_Fan_2019}, a system for automated game testing, utilizes deep reinforcement learning with the goal of discovering new bugs. This contrasts with BugCraft's objective of reproducing known bugs from user reports. Therefore, due to its focus on bug finding rather than reproduction, and its limitations like narrow action spaces and manual setup, Wuji is not a direct baseline for our work.

\subsection{Automated Bug Reproduction Using LLMs}

Automated bug reproduction entails converting unstructured bug reports into executable scripts reliably replicating the reported issues. LLMs have shown considerable success in generating code, particularly in creating unit tests \cite{Kang_Yoon_Yoo_2023} and enhancing bug report details \cite{Bo_Ji_Sun_Zhang_Wu_Wei_2024}. Recent frameworks leveraging LLMs have been developed to automate bug reproduction in mobile applications, particularly within Android environments, by interacting with the GUI \cite{Wang_Zhao_Feng_Zhang_Halfond}. These frameworks enable a series of structured actions where the LLM makes decisions based on the bug report, GUI interactions, and prior actions, while additional contextual information is provided as needed by the framework. Frameworks in \cite{Wang_Zhao_Feng_Zhang_Halfond} and \cite{Feng_Chen_2024} rely on methods that convert the GUI into textual representations, allowing the LLM to interact with and manipulate the interface through text-based commands. Additionally, a study \cite{Liu_Li_Chen_Wang_Wu_Wang_Hu_Wang_2024} incorporates visual inputs, enhancing the LLM's capacity to interpret and interact with complex GUIs. This approach has achieved performance comparable to other vision-based methods, highlighting the potential for LLMs to effectively combine textual and visual data in automated bug reproduction tasks.

\subsection{LLM Agents}
LLMs have shown great success at acting as artificial agents, which are entities capable of perceiving their surroundings, making decisions, and taking actions in response \cite{xi2023rise, yao2023react}. The agent framework involves equipping LLMs with tools, often in the form of APIs, to expand their action space and enable them to interact effectively with the environment. The most important part is that the LLMs were able to do this mainly by utilizing in-context learning \cite{tombrownfewshot}, alleviating the need to train specialized models for each in-domain task. Moreover, LLMs have also shown to be successful at taking action in an open-ended environment. This ability to act extends even to games like Minecraft, where the Voyager framework leveraged an LLM agent to achieve in-game progress milestones significantly faster than previous methods \cite{wang2023voyager}.

\section{Methodology} \label{sec:methodology}

We introduce BugCraft, a novel framework for end-to-end crash bug reproduction in complex interactive environments, using a custom dataset of Minecraft bug reports. As illustrated in Figure~\ref{fig:main_framework}, BugCraft operates in two phases: creating and executing S2R. The following sections detail our dataset (subsection~\ref{sec:dataprocessing}), BugCraft's components (subsections~\ref{sec:preprocessing}--\ref{sec:action_model}), implementation (subsection~\ref{sec:implementation_details}), and evaluation setup (Subsection~\ref{sec:study}).

\subsection{Data Curation} \label{sec:dataprocessing}

To our knowledge, no readily available dataset of Minecraft bug reports exists. Therefore, to quantitatively measure the efficiency of our framework, we constructed our own dataset of crash bug reports. Furthermore, we gathered data from the Minecraft Wiki to enrich our framework with in-game information. 

\subsubsection{BugCraft-Bench Dataset Curation} \label{sec:bugcraft-bench-curation}

We built a custom dataset consisting of 86 Minecraft: Java Edition crash bug reports to evaluate our framework. These reports were collected from Mojira, Minecraft's official bug tracker, utilizing Jira's REST API. We used a fine-grained query to retrieve all reports classified under the “Crash” category. Moreover, for each report, we gathered all associated comments, attachments, and external links (e.g., YouTube links), which may contain relevant files or videos. These were identified using regular expressions.

\begin{figure}[h]
    \centering
    \includegraphics[width=1\linewidth]{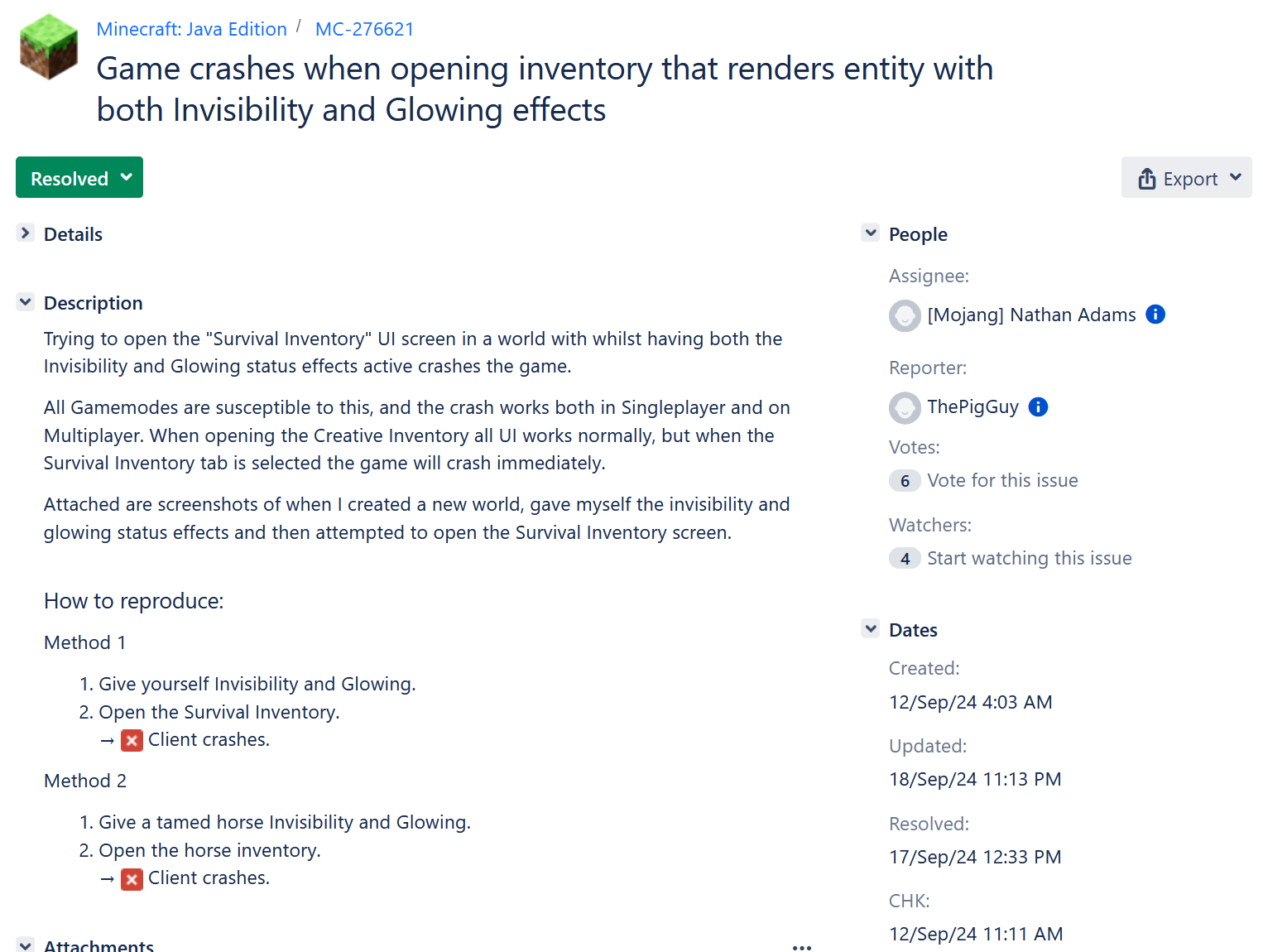}
    \caption{Bug Report Codenamed MC-276621}
    \label{fig:bug_report}
\end{figure}

In Figure~\ref{fig:bug_report}, we present a sample bug report, identified as MC-276621\footnote{\url{https://bugs.mojang.com/browse/MC-276621}}, which was verified and resolved by Mojang in a subsequent snapshot. This report illustrates a bug that can be reproduced through a specific combination of commands and UI interactions. Additionally, another user, contributed steps-to-reproduce (S2R), which were initially missing from the report. Importantly, the bug report (BR) in its final state lacks commands that would make reproducing the bug easier. Furthermore, though not relevant to this particular report, some bug reports, such as MC-275657\footnote{\url{https://bugs.mojang.com/browse/MC-275657}}, contain an implicit assumption that a version-specific setting is enabled without explicitly stating it, creating ambiguity during reproduction.

Each report in the dataset was categorized according to its "Confirmation Status" label in Mojira, which includes four possible states: "Confirmed," "Community Consensus," "Plausible," and "Unconfirmed." We primarily focused on "Confirmed" and "Community Consensus" reports, as these have been verified by the game's developer, Mojang Studios, making them reliable ground truth instances.

To ensure the quality and relevance of our dataset, we employed a multi-stage filtration process, focusing specifically on single-player crash bugs. We began by targeting the most recent bugs from the last five years (limited the size of the dataset because of resource constraints) crash bug reports from Mojira's database following our query results, spanning from the most recent, MC-277967\footnote{\url{https://bugs.mojang.com/browse/MC-277967}} (resolved on 05/Nov/24, affecting version 24w44a), to MC-145102\footnote{\url{https://bugs.mojang.com/browse/MC-145102}} (created on 19/Feb/19, affecting version 19w08b), covering a total of 70 different game versions. We then downloaded all associated data for each of these reports, including comments, attachments, and external links. Next, we undertook a manual review to categorize each report based on several criteria. Our focus was to isolate the single-player crash bugs, so we excluded reports involving multiplayer aspects, those requiring interactions between different game versions, and any that depended on external resources. We also removed reports related to issues outside the game itself and discarded reports if the final version differed so much from the initial draft that it could be considered a new submission. This review resulted in a refined set of 141 reports. Due to resource constraints and the significant cost associated with the LLM API calls, we chose to evaluate only a portion of the full dataset. As such, from the initial pool of 141 reports, we randomly selected a subset of 100 to create our final dataset. After our testing of the bug reports manually, we eliminated 14 of the bug reports due to being irreproducible, leaving us with a final number of 86 bug reports. This dataset was used for detailed analysis and the evaluation of our framework.

\subsubsection{Minecraft Wiki Dataset}

We use the official Minecraft Wiki\footnote{\url{https://minecraft.wiki}} site as our knowledge base. Since the base of the site used MediaWiki, we are able to get the wiki's text using the provided wiki API. We pulled every page in the wiki that was not a redirect page. We stored each page as a separate file for retrieval augmented generation (RAG). Using the wiki as our knowledge base allows us to circumvent reliance on our underlying LLM's pre-training data. 

\subsection{BugCraft Framework} \label{sec:bugcraft_framework}

\begin{figure*}[t] % Use [t] or [b] for full-width figures in two-column
    \centering
    \includegraphics[width=1\linewidth]{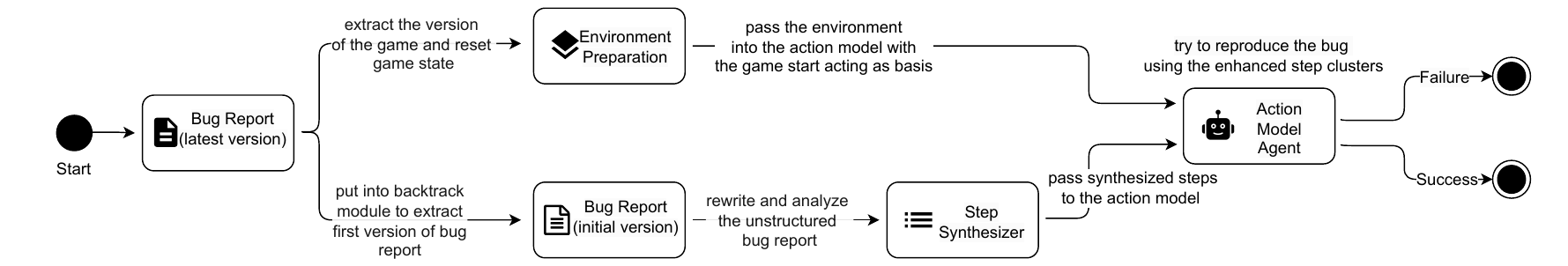}
    \caption{BugCraft Framework}
    \label{fig:main_framework}
\end{figure*}

The BugCraft framework (Figure~\ref{fig:main_framework}) is designed for automated bug reproduction and comprises two core components: the Step Synthesizer and the Action Model. These components work in tandem, with the Step Synthesizer generating a sequence of steps and the Action Model providing the executable actions, to transform raw, unstructured bug report data into executable scripts for accurate bug replication. We outline each component below.

\subsubsection{Preprocessing} \label{sec:preprocessing}
The initial stages of our process involve a crucial preprocessing phase applied to the raw bug report data. This phase begins with extracting the entirety of the textual content, encompassing the very first submission of the bug report and all subsequent comments. It's important to understand that we specifically focus on these initial versions because they represent the most challenging scenario for our framework. Specifically, we analyze the state of the bug report as it was initially submitted by the user before any modifications or refinements were made by the Mojira team. These initial reports are often less structured, potentially containing incomplete information or ambiguous language, reflecting the user's initial understanding and description of the issue. This contrasts significantly with later, revised versions, which benefit from the input and clarifications provided by experienced team members. The core challenge, and our framework's primary objective, is accurately interpreting and addressing these raw, initial reports. Subsequent refinements often simplify the report, making the issue easier to understand, but our focus is on the initial state of the report created by the user.

While the Mojira API lacks a direct method for retrieving previous versions, it does provide access to the issue's changelog, detailing the history of each field. We developed a back-tracker script to parse these fields and reconstruct the bug report at a specific point in its changelog. Comments are not included in the changelog, so we delete any comments that were added after the selected reconstruction step. To maximize the availability of information, the back-tracker stops at the latest state of the BR that has not been touched by a Mojira team member or a trusted individual (Users with [Helper] role). Using Mojira's role-based system, we specifically target the last version prior to any changes made by users with the roles [Mod], [Helper], or [Mojang].

\subsubsection{Step Synthesizer} \label{sec:step_synthesizer}

\begin{figure*}[t] % Use [t] or [b] for full-width figures
    \centering
    \includegraphics[width=1\linewidth]{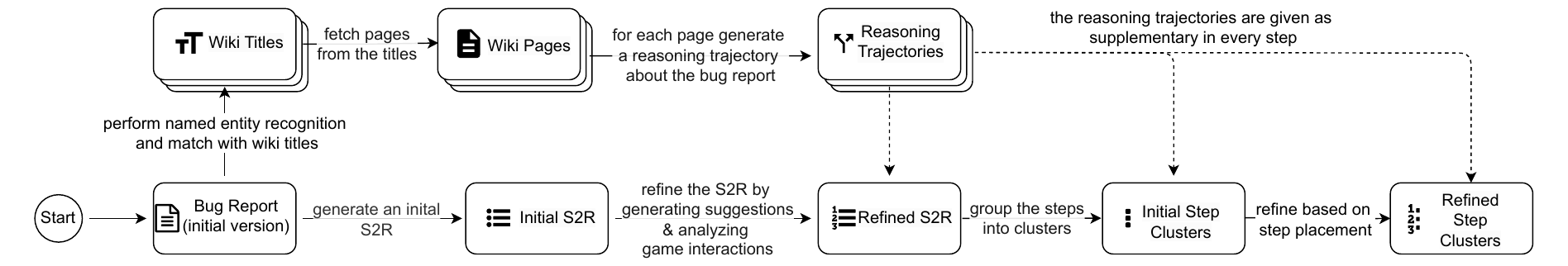}
    \caption{Step Synthesizer Component}
    \label{fig:step_synth}
\end{figure*}

As shown in Figure~\ref{fig:step_synth}, the Step Synthesizer component is specifically designed to handle the ambiguity and incompleteness inherent in many bug reports. Its core function is to transform the unstructured, and potentially unclear, textual descriptions into high-quality S2R steps. This process includes augmenting the steps with relevant external information to fill potential gaps. We employ a single VLM (Vision Large Language Model), referred to as the Step LLM, for the entire synthesizing process. The prompts used to guide the Step LLM are detailed in the reproduction package. This conversion happens through the following stages:

 \textbf{Knowledge Augmentation:} Prior to generating the S2R steps, we aim to identify relevant wiki pages to employ retrieval augmented generation (RAG) \cite{ragPaper}. This process begins by prompting the Step LLM with the bug report to perform named entity recognition, resulting in a list of entities. Fuzzy matching is then employed to compare this list against wiki page titles, yielding a set of relevant wiki page titles. Then, the Step LLM selects the most relevant titles from the set. In addition to these specifically identified pages, we incorporate generally relevant pages, such as those pertaining to the bug report's version, general game mechanics, and commands. Once all titles are selected, the corresponding pages are fetched, and individual analysis trajectories are generated for each page.  Inspired by the self-reasoning paper \cite{xia2024improvingretrievalaugmentedlanguage}, which generates a reasoning trajectory for each retrieved chunk, we adopt a similar approach to analyze each page. A reasoning trajectory is a step-by-step analysis generated by the Step LLM that examines the relevance of a specific external information source (e.g., a wiki page) to the bug report, highlighting key details and their potential implications for bug reproduction. This is crucial for handling the potential length of the pages and ensuring that only relevant information is cited. The LLM is also instructed to disregard any page deemed irrelevant. The resulting analysis trajectories, representing the relevant information extracted from the wiki pages, are appended to the end of every LLM call after S2R generation.

 \textbf{Initial S2R Generation:} Leveraging the bug report, comments, and web content, the Step LLM acts as a creative problem-solver to generate the initial S2R. It deconstructs the reporter's description to find the most reliable path to reproduction. A core part of this process is generating creative solutions that leverage Minecraft's command system to bypass difficult physical tasks. For example, instead of a step requiring the agent to physically navigate to a specific location and aim correctly to place a block, the synthesizer will generate a \texttt{/setblock} command. This strategic substitution is designed to minimize reliance on challenging spatial reasoning, thereby creating a clearer and more reproducible set of instructions that dramatically enhances the effectiveness of the Action Model.

\textbf{S2R Refinement:} Following the initial one-shot generation of the S2R, we subject it to a two-stage critique process, each stage utilizing a separate Step LLM call. This critique specifically targets inconsistencies within the plan and the predicted behaviors of in-game entities (mobs)\footnote{\url{https://minecraft.wiki/w/Mob}}. The initial S2R may contain errors regarding common mob behaviors, such as targeting, movement, and other related factors. To address this, we employ few-shot \cite{tombrownfewshot} examples to guide the Step LLM in generating suggestions for improvement. Subsequently, we utilize two additional Step LLM calls to rewrite the S2R, placing emphasis on incorporating the generated suggestions. This iterative refinement process \cite{madaan2023selfrefineiterativerefinementselffeedback} ensures the final S2R is comprehensive, internally consistent, and accurately reflects the expected behaviors of other agents in the game, such as mobs.

\begin{figure}[h] % IEEEtran often prefers [t] or [b]
    \centering
    \includegraphics[width=1\linewidth]{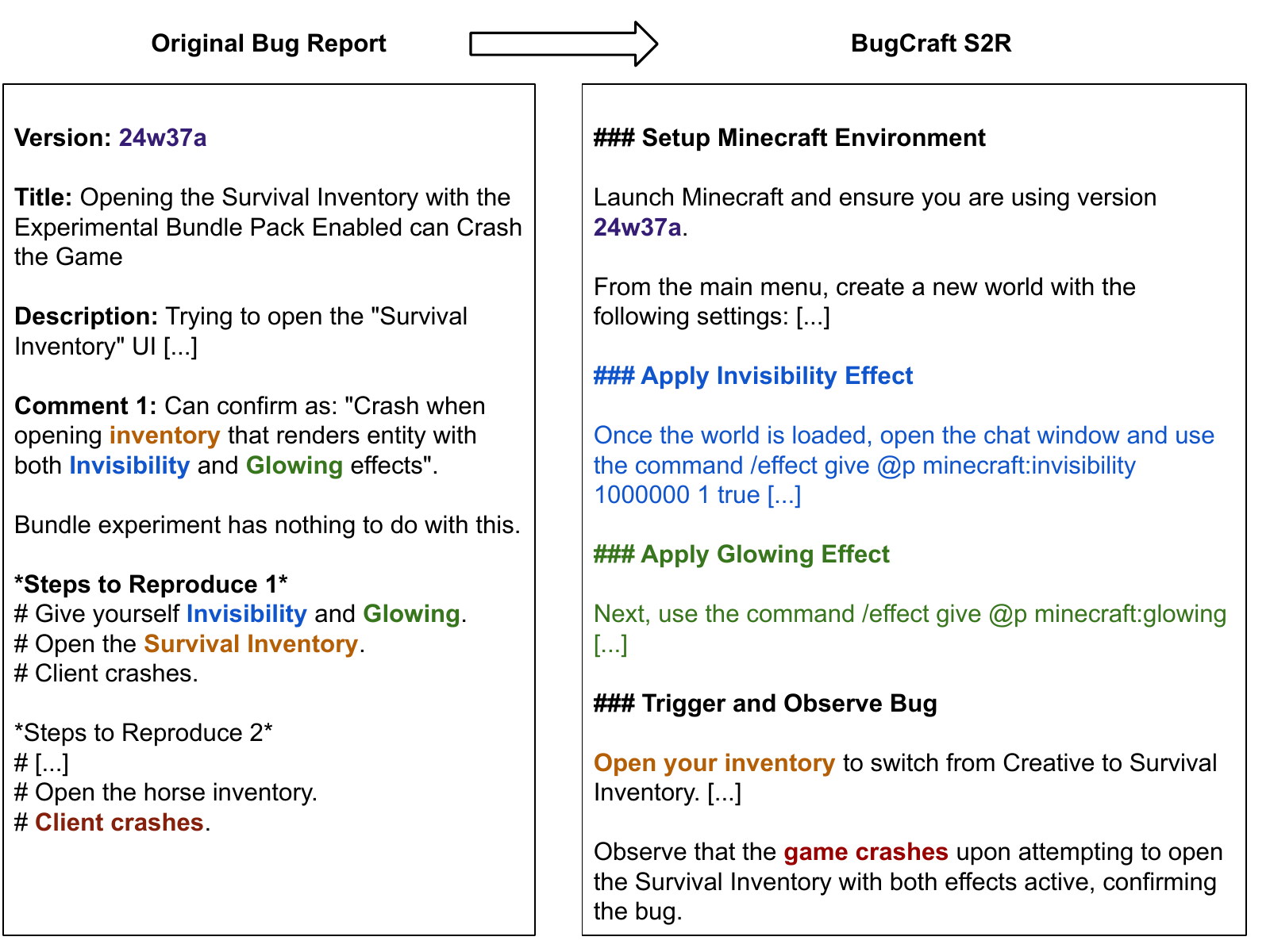}
    \caption{Initial MC-276621 Report and Our Output}
    \label{fig:bugcraft_example}
\end{figure}

\textbf{Step Clustering:} The finalized S2R is then used as input for a Step LLM prompt that instructs it to cluster the steps. This process empirically yields 2-4 clusters, each with a descriptive title and a list of its constituent steps. This clustering improves the readability and digestibility of the S2R. Following cluster generation, a critique is performed to identify any misplaced steps. Based on this critique, steps are reordered as needed. The outcome of this process is a refined and finalized S2R, optimized for clarity and downstream tasks. For instance, in MC-276621, our system effectively integrated the initial bug report and subsequent comment, enhancing the S2R with relevant commands for easier replication (Figure~\ref{fig:bugcraft_example}).

Upon completion, the Step Synthesizer yields a refined, fully annotated S2R set, ready for execution by the Action Model.

\subsubsection{Action Model} \label{sec:action_model}
\begin{figure*}[t] % Use [t] or [b] for full-width figures
    \centering
    \includegraphics[width=1\linewidth]{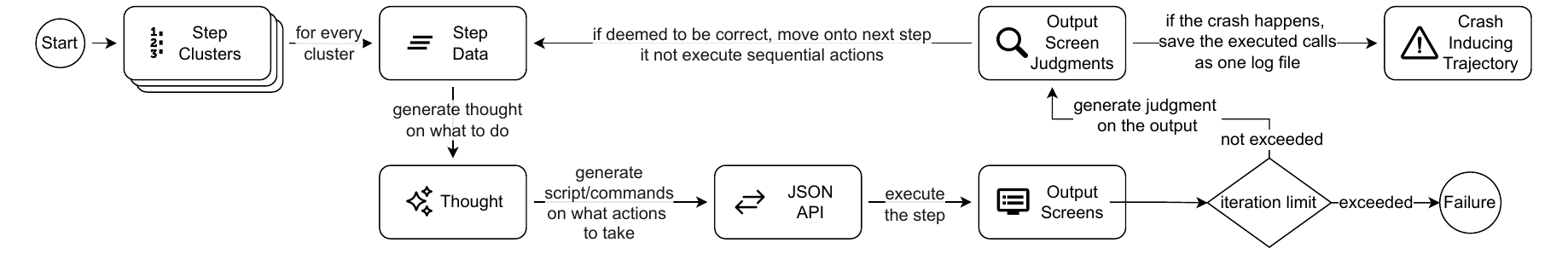}
    \caption{Action Model Component}
    \label{fig:action_model}
\end{figure*}

The Action Model component, illustrated in Figure~\ref{fig:action_model}, is responsible for translating the refined S2R into executable scripts. These scripts interact with the game interface, encompassing both the main menu and in-game elements. To achieve this, we developed a custom macro API, as existing solutions lacked a unified approach for handling both GUI and in-game interactions. Furthermore, the development of a custom macro API enabled the inclusion of game-specific methods tailored for frequently used interactions. To enhance the LLM's ability to identify and interact with specific GUI elements, we integrated OmniParser \cite{lu2024omniparserpurevisionbased}. OmniParser is a state-of-the-art open-weights model framework for parsing UI screenshots into structured elements. It excels at reliably identifying interactable icons and texts within the user interface and understanding the semantics of various elements in a screenshot while providing annotations.

\paragraph{Base Macro API}

The custom API uses a JSON-based REST schema, allowing the Action Model to interact with the game environment. The API supports the following methods:
\begin{itemize}
    \item \textbf{press(key: Key|List[Key], time: Optional[float]):} Handles keyboard and mouse input. If given multiple keys, it presses them simultaneously. The optional time parameter dictates the duration (in seconds) for which the keys are held down; if omitted, the keys are pressed and immediately released.
    \item \textbf{write(str: string):} Types the provided string into the currently focused text box using the keyboard.
    \item \textbf{command(instruction: String):} Executes Minecraft commands. Commands are restricted to non-menu contexts by the game engine. This is the shortcut for the LLM to open the chat, write the command, and then send the command message.
    \item \textbf{click(coordinates: Point):} Moves the mouse and clicks at the specified coordinate.
\end{itemize}
Each JSON request can encapsulate multiple methods that are executed sequentially. This allows for the orchestration of complex actions that involve menu navigation, gameplay commands, and GUI interactions. However, the click method is not directly used by the action-taking LLM. Instead, a \textbf{click\_place} method is provided. This method acts as an abstraction layer, bridging the OmniParser annotations with the API. The LLM specifies a particular bounding box identified by OmniParser, and the \textbf{click\_place} method translates this into a \textbf{click} command, using the center coordinates of the specified bounding box.

\paragraph{Action Model Framework}

Prior to initiating the logic loop, the game is launched using the version specified in the bug report, and the macro API is run in a separate thread. Finally, OmniParser is initialized and made ready to annotate images.

\begin{tcolorbox}[colback=black!5!white, colframe=black!75!black, 
title= Example Trajectory Iteration]
\textbf{Current Step Title:} Setup Minecraft Environment\\
\textbf{Thought:} We have accessed the 'More' options menu. The next step is to set the world type to 'Flatworld'. We need to click on 'World' to access the world type settings. The correct index for 'World' is 12.\\
\textbf{Action:} Clicked the place that had content: world. at coordinates: [x1: 0.41, y1: 0.06, x2: 0.60, y2: 0.12]

\textbf{Reflection:}
   The action was successful. The screen now shows the 'World' options menu, which includes the 'World Type' setting. This indicates that the 'World' button was clicked correctly, allowing access to the world type settings. This matches the expected outcome, as the next step is to set the world type to 'Flatworld'. 

\textbf{Classification:} SUCCESS
\end{tcolorbox}

For the Action Model loop, we implement a logic similar to ReAct \cite{yao2023react}. Further details regarding the prompts utilized in the Action Model are provided in the reproduction package. For the entire action-taking process, we use the same VLM, which we call the Action LLM. To execute each S2R, the Action Model loop is as follows:

\textbf{OmniParser Annotation:} Before each action, a screenshot of the Minecraft window is captured for annotation by OmniParser. OmniParser analyzes the image, identifying all icons and text elements.  Additionally, it classifies each element as either interactable or non-interactable. The resulting data is then formatted into a structured markdown table. %It performs OCR (Optical Character Recognition) on the text and employs a small VLM (Vision-Language Model) to annotate each icon.

\textbf{Thought \& Action Generation:} Prior to executing an action, the Action LLM is prompted to generate both a "thought," outlining its intended course of action, and the corresponding action itself in a single step. This prompt is supplied with the current trajectory (comprising previous thoughts, actions, and feedback), the annotated table generated by OmniParser, the corresponding game screenshot, and few shot examples to improve performance~\cite{tombrownfewshot}. The combination of visual input from the game image and the structured information from the annotation table facilitates the Action LLM's ability to identify and interact with specific game elements.

\textbf{Action Verification:} Following the generation of the thought and action, the Action LLM is prompted to verify whether the proposed thought and action are correct \cite{madaan2023selfrefineiterativerefinementselffeedback} and conducive to progressing the game state. This verification step acts as a crucial safeguard against errors and infinite loops, where the agent might repeatedly execute the same ineffective action. If the verification fails, the generated feedback, indicating the reason for failure, is used to prompt the generation of a new thought and action pair. This revised pair is then passed to the Macro API for execution.

\textbf{Feedback Generation:} After each action, a new screenshot is taken. The Action LLM then assesses the outcome, considering the prior thought and action. This involves judging if the thought was achieved and if the current step cluster is nearer completion, with a rationale. If all steps in the cluster are deemed complete, the Action LLM can signal a move to the next. This triggers a second Action LLM call to independently verify completion as a redundant check. If verification fails, the active cluster stays the same.

The action model loop runs until either the game crashes or a preset iteration limit is reached. Because the macro API logs each action, it generates a log capable of consistently reproducing the bug, assuming the bug isn't dependent on randomness. %Finally, the trajectory generated by the action model, if successful, is used to reiterate on the step clusters to generate the final S2R.

\subsubsection{Implementation Details} \label{sec:implementation_details}
We implement the framework using Python. We use LangChain\footnote{\url{https://www.langchain.com/}}, to implement the LLM calling logic. For the Action Model, we only keep the last 25 iterations in Action LLM's context to prevent context length overflow and set the iteration limit as 30 for our termination case. Moreover, while formatting the trajectory iterations, instead of putting the commands, we put verbal representations of what that command means. We do this to increase the interpretability of the generated actions for the LLM. 

For UI-Tars-1.5-7B, we adapt the implementation from OSWorld benchmark\cite{OSWorld} as it was done by the model's authors. We host the model on RunPod\footnote{\url{runpod.io}}, using VLLM\cite{kwon2023efficient}. For OpenAI-CUA, we again adapt OSWorld's implementation while following the official example on how to use OpenAI-CUA given by OpenAI\footnote{\url{https://github.com/openai/openai-cua-sample-app}} for guidance. We run both models using the inital BRs with 30 iterations. 

\subsection{Evaluation Setup} \label{sec:study}
To evaluate BugCraft, we pose three research questions about its components, considering real-world usage and comparative performance:\\
\textbf{RQ1:} How effective is our step synthesizer for producing correct plans?\\ 
\textbf{RQ2:} How effective is BugCraft at reproducing crash-inducing bug reports in an end-to-end manner?\\
\textbf{RQ3:} How does BugCraft compare to existing general GUI agents and specialized computer-use models?\\

% After cs491, preference benchmarkı olur bu:What is the preference of our generated bug reports from initial versions compared to final version bug reports?\\
%\textbf{RQ3:} What are the main strengths and weaknesses of the BugCraft framework?\\
\indent To address RQ1 and RQ2, we conducted experiments using a curated subset of our crash bug dataset, which we call the BugCraft-Bench. For RQ3, we evaluated BugCraft against UI-TARS-1.5-7B and OpenAI's Computer Use Agent on the same dataset. %This subset comprises 100 bug reports, carefully selected to represent a variety of scenarios that require interaction with both GUI and in-game world elements, as mentioned in section~\ref{sec:bugcraft-bench-curation}. The selection of this refined dataset was driven by the resource-intensive nature of running full evaluations, making it impractical to test all selected bug reports within the scope of this study.

\subsubsection{Evaluation of the Step Synthesizer} \label{sec:eval-of-step-synth}

To evaluate the accuracy of the step clusters generated by the Step Synthesizer, we manually reviewed each S2R. This involved attempting to manually reproduce the bug in Minecraft by strictly following the generated steps. If ambiguities or obstacles were encountered, we first consulted the original bug report for clarification. If the original report was insufficient, we referred to the final version of the bug report on Mojira and other relevant online resources (e.g., Minecraft Wiki) to understand the intended reproduction process. Each step cluster was then labeled as either "True", "Faulty", or "Irreproducible".
\begin{itemize}
    \item \textbf{True}: Following the step cluster leads to successful bug reproduction. Minor deviations or extra guidance steps for the LLMs are permitted.
    \item \textbf{Faulty}: The step cluster is flawed and prevents bug reproduction. It may be missing crucial steps, contain incorrect commands, or have logical inconsistencies.
    \item \textbf{Irreproducible}: The bug cannot be reproduced due to reasons unrelated to the Step Synthesizer's output. Examples include dependence on outdated versions, specific hardware, fundamental impossibility or the bug not causing a crash when reproduced.
\end{itemize}

To ensure the reliability of our labeling, particularly the distinction between "Faulty" and "Irreproducible," we conducted an inter-rater agreement analysis. Prior to the main labeling task, we jointly inspected the results of a test run to establish the labeling categories and refine our understanding of the criteria. Subsequently, we independently labeled the 100 bug reports in our run for the step synthesizer. Inter-rater agreement was measured using both percentage agreement and Cohen's Kappa \cite{McHugh_2012cohen}. The results of this analysis, including a confusion matrix, are presented in section  \ref{sec:interrater-agreement-step-synth}.

Faulty step clusters were further analyzed to identify the specific type of error. Each faulty step cluster was assigned at least one of the following error categories:
\begin{enumerate}
    \item \textbf{Wrong Command:} An S2R is categorized here if it includes commands that are not recognized or parseable by Minecraft. %This often occurs during the S2R enhancement phase, where the LLM may rely on its pre-trained data, which might not encompass newer command syntax, resulting in the substitution of older, invalid command formats.
    \item \textbf{Missing Step:} An S2R falls under this category when it lacks a critical step, rendering the subsequent steps implausible. %This issue was more prevalent with bug reports from newer Minecraft versions. This could be attributed to the increasing complexity of the game with each update and the training data cutoff date of GPT-4o (October 2023)\cite{openai2024gpt4ocard}, potentially leading to an incomplete understanding of the latest game mechanics.
    \item \textbf{Logic Error:} This category includes steps that are impossible, significantly hinder reproduction, or are internally contradictory within the plan itself. For example, simultaneously setting the time to day and spawning hostile mobs (which would burn in daylight), or attempting actions restricted by an open inventory screen.
    
\end{enumerate}

To ensure the reliability of our labeling, we conducted an inter-rater agreement analysis. Prior to the main labeling task, we jointly inspected the results of a test run to establish the labeling categories and refine our understanding of the criteria. Subsequently, we independently labeled the 100 bug reports in our main dataset. Inter-rater agreement was measured using both percentage agreement and Cohen's Kappa. The results of this analysis are presented in section ~\ref{sec:step-synth-analysis}.

\subsubsection{Evaluation of the Action Model} \label{sec:eval-of-action-model}

To evaluate the Action Model's ability to reproduce bugs, we conducted a comprehensive analysis of its performance on the subset of 86 bug reports that were deemed reproducible during the Step Synthesizer evaluation (Section \ref{sec:eval-of-step-synth}). Each reproducible report was processed end-to-end, and the Action Model's success or failure in reproducing the bug was recorded.

\subsubsection{Determining Action Model Success or Failure}
The initial assessment of the Action Model's performance focused on whether it successfully reproduced the bug or not. We classified success as instances where the game crashed as described. Failures occurred when no crash was observed. Due to the straightforward nature of this assessment, a formal inter-rater agreement was not deemed necessary.

\subsubsection{Failure Analysis and Inter-Rater Agreement}\label{sec:failureanalysis-interrater}

To understand the reasons behind failures, we performed a detailed examination of each unsuccessful attempt. This involved identifying recurring patterns and common failure modes, which led to the development of a structured categorization rubric. Just like we did with the Step Synthesizer rubric, we collaboratively analyzed the results of the same test run to define the failure categories, refine our understanding of the criteria, and establish a shared understanding of potential edge cases in applying the rubric. We also used our understanding of common agent mistakes to enhance the rubric. The results of this analysis, are presented in section \ref{sec:action-model-analysis}.
% To ensure the reliability of this failure categorization, we followed a similar inter-rater agreement procedure as used for the Step Synthesizer evaluation (Section \ref{sec:eval-of-step-synth}).
% Subsequently, the two raters (the authors) independently categorized the failures for the 100 bug reports in the sampled bug reports. Inter-rater agreement for the failure categorization was then measured using percentage agreement and Cohen's Kappa \cite{McHugh_2012cohen}.
%\paragraph{Action Model Grading Rubric} %\label{sec:action-model-grading}

\textbf{A. On Step Synthesizer Success}

\begin{enumerate}
    \item \textbf{Success:} The agent successfully reproduces the bug by following the provided plan.
    \item \textbf{Failure:} The agent fails to reproduce the bug.
    \begin{enumerate}
        \item \textbf{Agent Incapability:} The agent fails to reproduce the issue due to its own limitations, despite a correct plan and sufficient framework capabilities.
        \begin{enumerate}
            \item \textbf{Stuck in Loop:} The agent enters a repetitive loop, hindering progress. (e.g., Menu Loop, Command Loop, Death Loop).
            %\begin{enumerate}
             %   \item \textbf{Menu Loop:} The agent is trapped within the game's menu system, particularly the "Create World" screen in newer versions.
              %  \item \textbf{Command Loop:} The agent repeatedly attempts to execute the same incorrect or unparseable command.
               % \item \textbf{Death Loop:} The agent continuously dies due to in-game factors, such as hostile mobs, preventing it from acting.
                %\item \textbf{Other Loop:} Any other repetitive loop not covered above. 
            %\end{enumerate}
            \item \textbf{Poor Decision Making:} The agent's suboptimal choices lead to failure. This includes neglecting environmental factors, skipping crucial steps, or taking misguided actions, as detailed in the initial category descriptions.
        \end{enumerate}
        \item \textbf{Framework Incapability:} The framework's limitations make bug reproduction impossible or extremely difficult. This includes the inability to execute actions within tight time constraints or to interact with specific menu elements.
    \end{enumerate}
    \item \textbf{Error:} The action model encounters an internal error and crashes (e.g., the LLM failing to follow the JSON schema).
\end{enumerate}

\textbf{B. On Step Synthesizer Failure}

\begin{enumerate}
    \item \textbf{Recovery:} Despite a faulty plan from the Step Synthesizer, the agent successfully reproduces the bug.
    \item \textbf{Total Failure:} Both the Step Synthesizer and the Action Model fail to reproduce the bug.
    \item \textbf{Error:} The action model encounters an internal error and crashes (e.g., Pydantic error).
\end{enumerate}

%\noindent \textbf{Complementary Qualitative Data Collection:}
%To further contextualize our findings and validate the problem's significance from an industry perspective, we conducted a semi-structured interview, which was recorded, with a key decision maker of Minecraft (P1) on May 2025. P1 emphasized the critical importance of successful reproduction, noting that once a bug is reproducible, a significant portion of the diagnostic challenge is considered to be overcome, bringing considerable relief to the development team. This interview confirmed the substantial manual effort involved in current crash bug reproduction and highlighted the potential utility of automated solutions like BugCraft.

\section{Results} \label{sec:results}

This section evaluates BugCraft’s Step Synthesizer and Action Model end-to-end, then benchmarks baseline models and an additional model on our framework.

\subsection{Experimental Setup}

Our experiments were conducted on a high-performance machine running Windows 11, equipped with 32GB of RAM, an Intel i9-14900HX CPU, and an NVIDIA RTX 4090 Mobile GPU.
We used OpenAI's GPT-4o (specifically, the \texttt{gpt-4o-2024-11-20} snapshot) as the underlying language model for both the Step Synthesizer and Action Model components; we chose this model for its state-of-the-art performance and widespread adoption \cite{openai2024gpt4ocard}.
To demonstrate that our framework is model-agnostic we also evaluated the identical framework with GPT-4.1 (\texttt{gpt-4.1-2025-04-14}), keeping every other hyper-parameter unchanged.
All LLM calls in our framework were executed with temperature = 0 to ensure deterministic outputs \cite{renze2024effectsamplingtemperatureproblem}.
OmniParser served as our annotation backbone owing to its superior accuracy over comparable tools, and baselines (OpenAI CUA and UI-TARS-1.5-7B) are detailed in Section~\ref{sec:baseline-comparison}. %This powerful hardware, especially the GPU, was crucial for supporting the computationally intensive OmniParser component, which can utilize up to 12GB of VRAM during operation. This setup ensured that our evaluations were not limited by computational resources, allowing us to thoroughly assess BugCraft's capabilities under demanding conditions.
% bu kısmı direk kapa paperı sadece kötü gösteriyor
% The complete testing process for the 100 bug reports took approximately 36 hours. This duration is attributed to the inherently sequential nature of testing each bug report, as parallelization was not feasible. Individual runs could take up to 30 minutes, depending on the complexity of the bug and the actions required for reproduction.\footnote{\url{https://minecraft.wiki/w/Java_Edition_19w13b}}

\subsection{Step Synthesizer Analysis} \label{sec:step-synth-analysis}
\begin{table}[h] % IEEEtran often prefers [t] or [b]
    \centering
    \caption{Step Synthesizer Planning Correction Analysis}
    \label{tab:planning_correction}
    \begin{tabular}{l|c|c}
        \hline
        \textbf{Outcome} & \textbf{Count} & \textbf{Percentage (\%)} \\
        \hline
        True & 57 & 66.28\% \\
        Faulty & 29 & 33.72\% \\
        \hline
        \noalign{\vspace{1.5pt}} % Add vertical space here
        \multicolumn{3}{l}{\textbf{Faulty Analysis Breakdown}\textsuperscript{\textbf{*}}:} \\
        \hline
        \hspace{0.5cm} Wrong Command & 14 & 48.28\% \\
        \hspace{0.5cm} Missing Step & 10 & 34.49\% \\
        \hspace{0.5cm} Logic Error & 9 & 31.03\% \\
        \hline
        Irreproducible  & 14 & - \\
        \hline
        \noalign{\vspace{1.5pt}} % Add vertical space here
        \multicolumn{3}{l}{\textsuperscript{\textbf{*}}\footnotesize{One faulty case may belong to multiple categories.}}
    \end{tabular}
\end{table}

Table~\ref{tab:planning_correction} summarizes the Step Synthesizer using GPT-4o evaluation (Section~\ref{sec:eval-of-step-synth}), where step clusters were manually tested for bug reproduction and categorized. The findings indicate 66.28\% of clusters were "True" (successful reproduction), while 33.72\% were "Faulty" (preventing reproduction). The primary reasons for faulty plans were "Wrong Command" (48.28\%), "Missing Step" (34.49\%), and "Logic Error" (31.03\%). After our thorough manual analysis, 14 reports were classified as "Irreproducible" due to factors like hardware issues or original report errors, not synthesizer faults. %These results suggest that while the Step Synthesizer is generally effective in creating reproduction plans, it still faces challenges in generating correct Minecraft commands, ensuring all necessary steps are included, and maintaining logical consistency within the plan.

\subsubsection{Inter-Rater Agreement for Step Synthesizer} \label{sec:interrater-agreement-step-synth}
\begin{figure}[h] % IEEEtran often prefers [t] or [b]
    \centering
    % \includesvg has issues with some LaTeX distributions/compilers.
    % It's safer to convert SVG to PDF and use \includegraphics.
    % Assuming figures/reproducibility_confusion_matrix.pdf exists
    \includegraphics[width=0.4\textwidth]{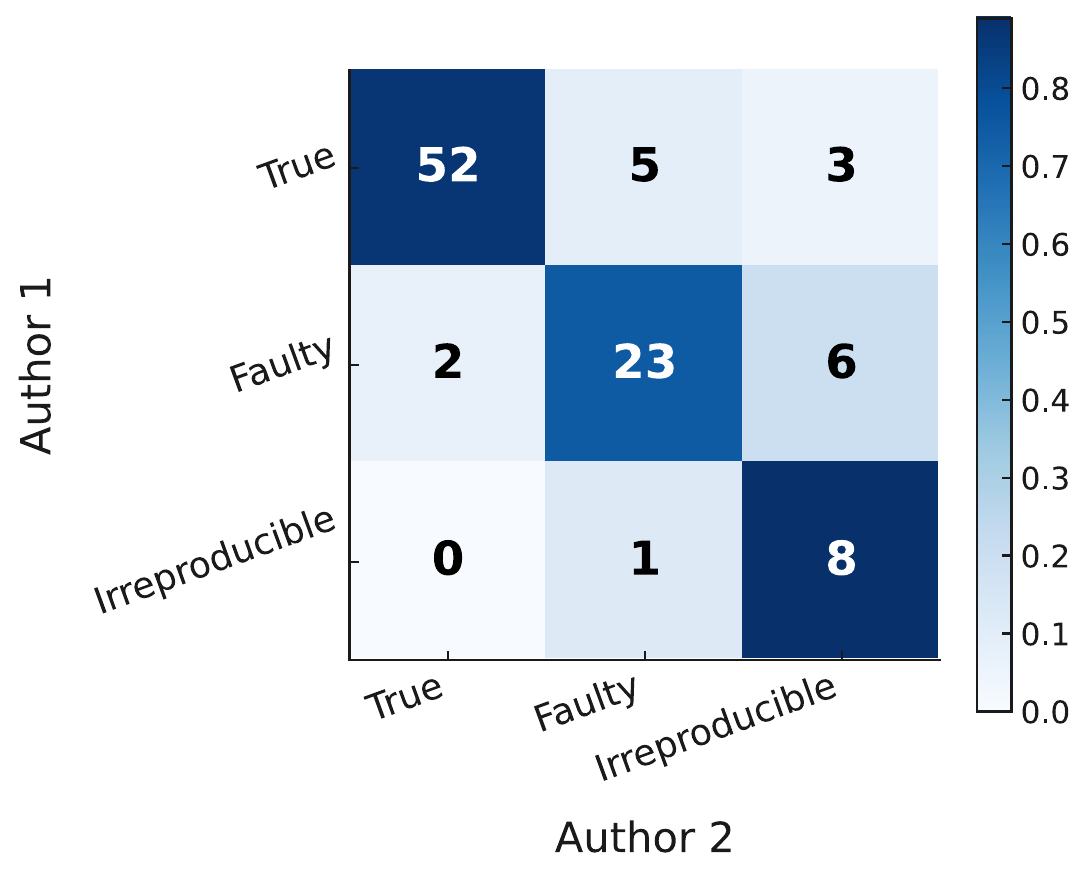}
    \caption{Confusion matrix of bug report reproducibility showing inter-rater agreement}
    \label{fig:confusion-matrix}
\end{figure}

To ensure the reliability of our manual labeling of step clusters, we conducted an inter-rater agreement analysis. The results are presented in Figure~\ref{fig:confusion-matrix}. The confusion matrix shows the agreement and disagreement between the two authors on the classification of step clusters as "True," "Faulty," or "Irreproducible." A Cohen's Kappa of 0.70 and a percentage agreement of 83.0\% were achieved, indicating substantial agreement.

%Furthermore, we analyzed the inter-rater agreement on the specific reasons for failure when a step cluster was labeled as "Faulty". The confusion matrix in Figure~\ref{fig:confusion-matrix} displays this agreement. The categories considered were "Wrong Command," "Missing Step," and "Logic Error." We obtained a Cohen's Kappa of 0.70 and a percentage agreement of 83\%, which also indicates a good level of agreement on the specific types of errors.

\subsection{Action Model Analysis} \label{sec:action-model-analysis}

\begin{table}[h] % IEEEtran often prefers [t] or [b]
\centering
\begin{threeparttable} % Start threeparttable
\caption{Experimental Results and Analysis of Failure Cases}
\label{tab:results}
\begin{tabular}{l|c|c}
\hline
\hline
\textbf{Category} & \textbf{Count} & \textbf{Percentage (\%)} \\
\hline
\hline
\textbf{A. Step Synthesizer Success} & 57 & 100.0\% \\
\quad Success (Bug Reproduction) & 22 & 38.60\% \\
\quad Failure & 32 & 56.14\% \\
\quad \textit{\textbf{Breakdown of Failures:}} & & \\
\quad $\circ$ Agent Incapability & 29 & 50.88\% \\
\quad \quad - Stuck in Loop & 11 & 19.30\% \\
\quad \quad \quad $\bullet$ Menu Loop & 8 & 14.04\% \\
\quad \quad \quad $\bullet$ Command Loop & 2 & 3.51\% \\
\quad \quad \quad $\bullet$ Death Loop & 1 & 1.75\% \\
\quad \quad - Poor Decision Making & 17 & 29.82\% \\
\quad $\circ$ Framework Incapability & 4 & 7.02\% \\
\quad Error & 3 & 5.26\% \\
\hline

\textbf{B. Step Synthesizer Failure} & 29 & 100.0\% \\
\quad Recovery & 4 & 13.79\% \\
\quad Failure & 24 & 82.76\% \\
\quad Error & 1 & 3.45\% \\
\hline
\hline
\textbf{Overall Totals} & & \\
\quad Total Trials & 86 & 100.0\% \\
\quad Total Success & 26 & 30.23\% \\
\quad Total Failure & 56 & 65.12\% \\
\quad Total Errors\tnote{*} & 4 & 4.65\% \\
\hline
\hline
\end{tabular}
\begin{tablenotes}
    \footnotesize
    \item[*] Three errors were from internal crashes (LLM failed to give the output in correct JSON schema) and one error was from the technical issues that the potential bug could cause.
\end{tablenotes}
\end{threeparttable} % End threeparttable
\end{table}

Table~\ref{tab:results} summarizes the Action Model's performance using GPT-4o on 86 bug reports, following the methodology in section~\ref{sec:failureanalysis-interrater}. Initially, 100 reports were considered; however, 14 were deemed "Irreproducible" (Section~\ref{sec:step-synth-analysis}), and 4 Action Model runs encountered errors (3 internal, 1 bug-induced technical issue), leading to 86 attempted reproductions. These included cases with faulty step clusters, where the agent could potentially recover.

In comprehensive end-to-end testing, the Action Model successfully reproduced 26 out of 86 (30.2\%) with GPT-4o. When provided with correct step clusters ("Step Synthesizer Success"), the success rate was 38.60\% (22 out of 57). Remarkably, when step clusters were faulty ("Step Synthesizer Failure"), the Action Model still recovered and reproduced the bug in 4 out of 29 cases (13.79\%), showcasing the agent's ability to overcome plan errors. Failures with correct plans were primarily due to "Agent Incapability" (50.88\%), encompassing "Poor Decision Making" (29.82\%) and "Stuck in Loop" (19.30\%). "Framework Incapability" contributed to 7.02\% of these failures. In scenarios where the Step Synthesizer provided faulty plans and the Action Model did not recover, the overwhelming majority of failures (82.76\%) were attributed to these incorrect steps.

\subsection{Comparative Evaluation with Baselines} \label{sec:baseline-comparison}

We evaluated BugCraft against two competitive baselines: OpenAI's Computer Use Agent (CUA)\cite{openai2025cuacard}, a recent commercial computer-use model, and UI-TARS-1.5-7B, a strong open-source GUI automation model. All systems were tested on the identical 86 bug reports without any errors from BugCraft-Bench to ensure fair comparison. We use OSWorld's hyperparameters\cite{OSWorld} when running the benchmark.

\begin{table}[h]
\centering
\caption{Baseline Comparison: Performance and Cost Analysis}
\label{tab:baseline-comparison}
\begin{tabular}{l|c|c|c|c}
\hline
\textbf{System} & \textbf{Success} & \textbf{Active} & \textbf{MTTR} & \textbf{Cost/} \\
                & \textbf{Rate} & \textbf{Time} & & \textbf{Attempt} \\
\hline
Human Expert\footnotemark & ~83\% & 20 min & 3.41 days & \$28.20 \\
\hline
BugCraft (GPT-4.1) & \textbf{34.9\% (30)} & 10.00 min & 10.00 min & \$1.16 \\
BugCraft (GPT-4o)& 30.2\% (26) & 15.56 min & 15.56 min & \$1.45 \\
OpenAI CUA & 25.5\% (22) & 6.37 min & 6.37 min & \$0.65 \\
UI-TARS-1.5-7B & 0.0\% (0) & \textbf{3.27 min} & \textbf{3.27 min} & \textbf{\$0.02} \\
\hline
\multicolumn{5}{l}{}
\end{tabular}
\end{table}
\footnotetext{Active time, the manual effort required to reproduce a bug, measured by the avg. time authors spent during dataset validation. MTTR 
(Mean Time To Reproduce) is the avg. time for an 
issue confirmation, calculated from Mojira API. Cost is 
estimated at \$85/hour (\$64 median wage + 30\% employer costs 
\cite{bls2025ecec}). Success rate is calculated from our percentage-agreement analysis.}

The experimental results are presented in Table~\ref{tab:baseline-comparison}. BugCraft with GPT-4.1 achieved a 34.9\% success rate, outperforming OpenAI CUA (25.5\%) and UI-TARS-1.5-7B (0\%). The performance improved from 30.2\% with GPT-4o to 34.9\% with GPT-4.1. Regarding execution efficiency, BugCraft completed reproductions in 10.00 minutes on average with GPT-4.1 (15.56 minutes with GPT-4o), showing a trade-off between success rate and speed compared to CUA's 6.37 minutes and UI-TARS's 3.27 minutes. The cost analysis shows that BugCraft operates at \$1.16 per attempt (GPT-4.1), placing it between the more economical CUA (\$0.65) and the low-cost UI-TARS (\$0.02), which reflects the computational demands of its two-stage approach. Notably, while human experts achieve ~83\% success rate with only 20 minutes of active work, their mean time to reproduce (MTTR) extends to more than this time due to organizational processes, at \$28.20 per attempt based on industry-standard developer costs.

\section{Discussion} \label{sec:discussion}
This study evaluated the BugCraft framework via three research questions: step-synthesizer planning (RQ1), end-to-end reproduction (RQ2), and baseline comparison (RQ3).

\subsection{Addressing RQ1}

The Step Synthesizer with GPT-4o demonstrated a promising ability to transform raw, unstructured bug report data into structured, actionable steps for bug reproduction. It achieved a 66.28\% success rate in generating accurate plans (step clusters), as evaluated through the methodology outlined in section~\ref{sec:eval-of-step-synth}. This indicates that the component is generally effective in understanding the reported issue, extracting relevant information, and synthesizing it into a coherent sequence of actions. The strong correlation between successful plan generation and the Action Model's success further validates the Step Synthesizer's performance, highlighting its crucial role in the overall bug reproduction pipeline. However, the 33.72\% failure rate reveals that challenges remain. The analysis of faulty plans identified three main error categories: "Wrong Command" (48.28\%), "Missing Step" (34.49\%), and "Logic Error" (31.03\%). "Wrong Command" errors often stemmed from the LLM relying on outdated command syntax from its pre-training data, especially for newer Minecraft versions. This was particularly evident in the LLM's struggle to generate correct item tags, which have a detailed schema that changes frequently between versions\footnote{\url{https://minecraft.wiki/w/Data_component_format}}. Additionally, some hallucinated commands were generated that never existed in the game. "Missing Step" errors indicated that the Step Synthesizer sometimes failed to capture all the crucial details needed for successful reproduction, particularly for complex bugs or those involving recent game mechanics. This often manifested as the LLM failing to adequately track and account for the evolving game state. "Logic Error" cases demonstrated challenges in ensuring the generated plan's internal consistency and adherence to game rules. The Step Synthesizer offers a good foundation for bug reproduction planning. Integrating version-specific game information, command verification, and logical consistency checks could further improve its reliability. Solutions could include enhancing external data integration and incorporating game environment feedback to refine plans dynamically, however these are challenging to pull off due to their necessity to work in many versions of the game.

\subsection{Addressing RQ2}

BugCraft using GPT-4o achieved a 30.2\% success rate in end-to-end reproduction of crash-inducing bug reports from our curated dataset. This demonstrates a tangible capability for automated game bug reproduction, showcasing the potential of LLMs to bridge the gap between natural language descriptions of bugs and concrete actions within a complex game environment. The primary bottleneck in our end-to-end pipeline was the Action Model. Even when provided with accurate step-by-step plans from the Step Synthesizer, the Action Model still failed to reproduce the bug in a majority of cases (56.14\%). This is an expected result as even humans can sometimes struggle at reproducing these bugs given a well defined plan. Agent incapability was the most significant factor hindering the Action Model's performance. This category encompassed both "Poor Decision Making" (29.82\%) and being "Stuck in Loops" (19.30\%). "Poor Decision Making" highlighted the challenges faced by the Action LLM in navigating the complex, visually-rich Minecraft environment. Operating solely on visual input, without a deeper symbolic understanding of the game state, the LLM sometimes struggled to make optimal choices, overlooking environmental cues or deviating from the intended plan. The "Stuck in Loop" failures further illustrated these challenges, manifesting as "Menu Loops" (14.04\%), "Command Loops" (3.51\%), and even a "Death Loop" (1.75\%). "Menu Loops," particularly within the "Create World" screen, indicated difficulties in reliably navigating complex GUI structures. "Command Loops" arose from issues in command parsing or execution, with the agent repeatedly attempting unparseable or ineffective commands. The single "Death Loop" instance was caused by hostile mobs continuously killing the player, preventing progress.

Beyond agent-specific limitations, "Framework Incapability" contributed to 7.02\% of failures when correct steps were available. This suggests that while our custom API provided a foundational level of interaction, it lacked the granularity or robustness needed for certain bug reproduction scenarios. Limitations in precise mouse control, rapid action execution, or interaction with specific UI elements within the game hindered the Action Model's ability. Notably, the Action Model successfully reproduced bugs in 13.79\% of cases despite faulty plans from the Step Synthesizer. These "Recovery" cases suggest the agent can explore the given S2R and make concious decisions on the fly to correct the wrong steps. While BugCraft shows promise for automated game bug reproduction, improvements in agent capabilities and framework robustness are essential. Future work should focus on enhancing the agent's understanding of the game environment, possibly via symbolic or hybrid approaches. Improving the game interaction API for better control/feedback, and adding more robust error detection and recovery, could also boost performance.

\subsection{Addressing RQ3}

\subsubsection{Isolation of Framework Contributions}

Testing against general GUI agents (UI-TARS: 0\% success) and specialized computer-use models (OpenAI CUA: 25.5\% success) isolates BugCraft's contributions. The 36.9\% success rate, a 32\% relative improvement over CUA, stems from three components: (1) Step Synthesizer with game-specific RAG, (2) flexible step clustering, and (3) unified game interaction APIs. These specifically address what generic approaches miss: game domain knowledge and spatial reasoning circumvention.

\subsubsection{Performance Justification and Efficiency Analysis}

While BugCraft's 34.9\% success rate is lower than the 83\% achieved by human experts, this overlooks crucial operational advantages. The system’s true value is revealed in its superior efficiency, cost-effectiveness, and scalability. BugCraft demonstrates remarkable time efficiency, completing a reproduction in 10 minutes compared to a human's 3.41 day turnaround, which is often delayed by organizational overhead. This vast reduction in time allows our system to be suitable for rapid triage. Financially, the system offers a 10-fold cost reduction per successful reproduction-\$3.44 for BugCraft versus \$35.25 for a human - and its success with crash bugs is definitively verifiable. Finally, unlike human experts who work sequentially, the framework processes bugs in parallel across multiple instances. This capability is critical for projects like Minecraft that handle hundreds of bug reports weekly, allowing for a high-throughput, automated workflow that humans cannot match.

\subsubsection{Technical Differentiation from Prior Work}

BugCraft differs from prior mobile/web GUI automation through three game-specific innovations: (1) intelligent bug analysis by Step Synthesizer, (2) RAG integration with game knowledge bases to disambiguate vague reports, and (3) unified macro API design handling both 2D menus and 3D gameplay. These address challenges unique to games, spatial reasoning, complex state dependencies, and version-specific mechanics, that 2D GUI frameworks cannot handle.

\begin{figure}[t] % Use [t] or [b] for full-width figures
    \centering
    \includegraphics[width=1\linewidth]{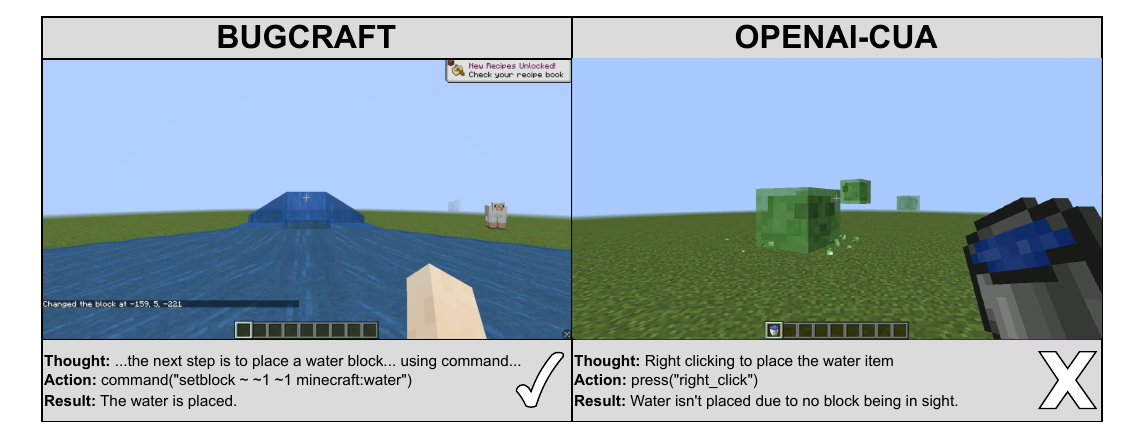}
    \caption{Case study analysis of MC-161902 comparing how our framework successfully places the water block whereas OpenAI-CUA fails.}
   \label{fig:case_study}
\end{figure}
\textbf{Case Study:} We analyzed bug MC-161902\footnote{\url{https://bugs.mojang.com/browse/MC-161902}}, a game crash triggered when specific mobs contact water, to highlight a critical difference between BugCraft and baseline models. The core challenge to reproduce this bug is placing the water as shown in Figure~\ref{fig:case_study}.
\begin{enumerate}
     \item \textbf{OpenAI CUA:} Despite being encouraged to use commands, the model opts for a physical approach. It successfully gets a water bucket but fails when trying to aim and place the water on a valid surface, as it is unable to see the difference and fails to reproduce the bug.

    \item \textbf{BugCraft:} Our Step Synthesizer analyzes the bug report and identifies the most efficient path to the goal. It intelligently bypasses the difficult physical task by using a command to spawn the water block directly. The Action Model then easily follows the steps and is able to reproduce the bug.
\end{enumerate}

\subsubsection{Qualitative Analysis of the Models}

OpenAI-CUA navigates menus better than our Action Model, but its poorer solution choices reveal the value of our Step Synthesizer’s bug analysis. UI-Tars-1.5-7B fails outright; as a very small model specialized for computer GUI usage, it does not know Minecraft well and gets lost in the menus or idles in-game. The authors also point out the fact that the smaller model is not optimized for game specific scenarios\footnote{\url{https://github.com/bytedance/UI-TARS\#limitations}}.
GPT-4.1 achieved a higher end-to-end success rate than GPT-4o (34.9\% vs. 30.2\%; 30 vs. 26 of 86). A qualitative execution-log analysis indicates GPT-4.1 followed the prescribed step clusters more closely, whereas GPT-4o exhibited more off-plan attempts. The models’ successes only partially overlapped (20 both; 10 only GPT-4.1; 6 only GPT-4o; McNemar p=0.45), yielding an oracle coverage of 36/86 (41.9\%). This complementarity implies that mixing agent strategies can increase coverage in bug reproduction.

\section{Threats to Validity} \label{sec:threats_to_validity}

\subsection{Construct Validity}
A potential threat to construct validity stems from using the Minecraft Wiki for knowledge augmentation. While BugCraft might generalize to other games, especially those with developer support, the exceptionally detailed Minecraft Wiki may not be representative of typical game documentation. This, combined with Minecraft's popularity, likely led to substantial Minecraft data in the LLM's pre-training. Games with less documentation or popularity might see reduced LLM effectiveness. To address this, we allow the Step Synthesizer to be augmented with any kind of external knowledge. To directly address the concern that LLMs might have been trained on final, corrected versions of Mojira bug reports, we made a conscious decision to use only the initial, unedited versions. We used the smaller UI-Tars-1.5-7B model, as the larger one was not publicly available. Despite following the official implementation, since small models are sensitive to prompt changes, our results may be suboptimal as we lacked access to potentially better internal prompt templates.

\subsection{Internal Validity}
For internal validity, we considered the inherent stochasticity of LLMs. To promote consistency and deterministic behavior in most components, we mitigated this by setting the temperature parameter to zero. While this is an unavoidable aspect of the game environment and a challenge for vision-based agents, we acknowledge that resource and time constraints limited our ability to conduct multiple runs for every bug report to fully quantify its impact in this study. Furthermore, the limited number of evaluators in the manual analysis phases could introduce bias. %A larger, more diverse group would have strengthened the assessment's reliability.

\subsection{External Validity}
A key consideration for external validity is that our evaluation is limited to crash bugs within Minecraft: Java Edition, which raises questions about direct generalizability to other bug types or different games. However, it is worth noting that the size of our evaluation dataset is comparable to automated bug reproduction frameworks in other domains-for instance, similar Android bug reproduction frameworks have been evaluated on datasets sized similarly to ours \cite{Wang_Zhao_Feng_Zhang_Halfond , Feng_Chen_2024}. Moreover, BugCraft demonstrated its adaptability by successfully processing reports from over 70 different Minecraft versions, spanning more than six years of the game's development. Minecraft itself, with its vast and evolving environment, provides a robust test case for our approach due to its inherent complexity. While these results within Minecraft are promising, caution is still warranted when extrapolating directly to all other bug types or game genres, highlighting the need for future research to explore BugCraft's performance in diverse contexts. Additionally, our reliance on publicly available Mojira bug reports, while transparent, might not fully represent the entire spectrum of issues encountered by all players.

%\section{Future Work} \label{sec:future_work} % Section typically not numbered like this for Conclusion
%Future work should address the challenges of agent development across numerous game versions, as our vision-based agent was a general solution. Collaboration with game developers could provide specialized APIs for enhanced in-game movement and interaction, similar to Voyager's high-level API, improving bug reproduction. Furthermore, comprehensive datasets and robust ground truth verification are crucial, especially for non-crash bugs in Minecraft. Future benchmarks should incorporate automated or semi-automated ground truth methods to evaluate a wider range of bug types. Assessing BugCraft's generalizability beyond Minecraft is also essential, requiring investigation into its adaptability to diverse game engines, genres, and complexities through comparative studies across multiple games. Our roadmap includes improving BugCraft based on identified weaknesses and extending the agent to handle multiplayer bugs and bugs involving file loading (Worlds, datapacks, world setting JSONs).

\section{Conclusion} \label{sec:conclusion} % Section typically not numbered like this for Conclusion
The paper introduces the BugCraft-Bench, a novel benchmark for Minecraft bug reproduction, and BugCraft, an LLM-based framework for end-to-end crash bug reproduction in Minecraft. BugCraft-Bench addresses the dataset gap in game bug reproduction research, offering a curated set of 86 Minecraft crash bug reports for standardized evaluation. BugCraft framework achieves a 34.9\% end-to-end success rate in reproducing crash bugs, outperforming existing computer-use approaches, highlighting LLMs' potential for automating game bug reproduction. However, limitations in the Action Model, such as the agent getting stuck in loops and poor decision making by the agent were identified, along with Step Synthesizer challenges in command accuracy and detail completeness, especially for newer Minecraft versions.

Moving forward, several key areas warrant further investigation. Future work should focus on enhancing LLM game environment understanding, potentially through collaboration with game developers to create specialized APIs for enhanced in-game movement and interaction, similar to the high-level API Voyager \cite{wang2023voyager} uses, called Mineflayer \cite{mineflayer}. This could also improve bug reproduction across numerous game versions, addressing the limitations of our current vision-only based approach. Developing comprehensive datasets and robust ground truth verification methods, especially for non-crash bugs, is crucial, requiring future benchmarks to incorporate automated or semi-automated verification tools. Finally, assessing BugCraft's generalizability is essential, necessitating comparative studies across diverse game engines, genres, and complexities. 
Our roadmap focuses on improving BugCraft’s Action Model and Step Synthesizer based on identified weaknesses, plus adding support for multiplayer and file-loading bugs (worlds, datapacks, JSON settings).
 Ultimately, the advancements presented in this paper provide a strong foundation for future advancements in automated bug reproduction, paving the way for more automatic game development processes.
\newpage
% --- IEEE prefers references to be in a separate .bib file ---
\bibliographystyle{IEEEtran}
\bibliography{references} % Assuming your .bib file is named references.bib

\end{document}